# Upper Critical Field in the Molecular Organic Superconductor (DMET)$_2$I$_3$

Pashupati Dhakal,[1,*] Harukazu Yoshino,[2] Jeong Il Oh,[1] Koichi Kikuchi[3]

and Michael J. Naughton[1,+]

[1] Department of Physics, Boston College, Chestnut Hill, MA 02467 USA

[2] Graduate School of Science, Osaka City University, Osaka 558-8585 Japan

[3] Graduate School of Science and Engineering, Tokyo Metropolitan University, Tokyo 192-0397

Japan

Abstract

We report the temperature dependence of the upper critical magnetic field in the quasi-one-dimensional molecular organic superconductor (DMET)$_2$I$_3$, for magnetic field applied along the intrachain, interchain, and interplane directions. The upper critical field tends to saturation at low temperature for field in all directions and does not exceed the Pauli paramagnetic limit. Superconductivity in (DMET)$_2$I$_3$ thus appears to be conventional spin singlet, in contrast to the status of the isostructural Bechgaard salts. We also discuss a magnetic field-induced dimensional crossover effect in the normal metallic state which had previously appeared to be associated with superconductivity.

## I. INTRODUCTION

After the discovery of superconductivity in Bechgaard salt (TMTSF)$_2$PF$_6$ by Jerome *et al.* [1] in 1980, there has been sustained interest in the nature and origin of superconductivity in low-dimensional molecular organic superconductors [2]. This is because not only is it remarkable that such quasi-one dimensional (Q1D) organic materials superconduct, let alone conduct, but aspects of the superconductivity, which competes with spin density wave (SDW) antiferromagnetism in this system, were found to be anomalous. Among these was the behavior of the upper critical field, especially for field oriented in the highly conducting *x-y* plane (formed by a sheet of highly conducting 1D chains along *x*). In particular, $H_{c2}^y$ (*H* in-plane, perpendicular to the chains) showed pronounced upward curvature with no sign of saturation, eventually even exceeding $H_{c2}^x$ [3]. Moreover, at low temperature, $H_{c2}^y$ for (TMTSF)$_2$ClO$_4$ [4,5] and (TMTSF)$_2$PF$_6$ [3,6] was shown to significantly exceed the Pauli paramagnetic limit ($H_p$) imposed by quantum statistics on spin singlet superconductors. These unusual high critical fields, as well as NMR relaxation [7,8] and Knight shift [9] results, among others, suggested the possibility that equal spin triplet pairing was responsible for the superconductivity. Recent NMR results in lower magnetic fields, however (~0.9T compared to 1.4T), favor the existence of line nodes on the Fermi surface, which the authors interpret as suggestive of *d*-wave, spin singlet pairing [9,10], at least in the field regime measured. However, *d*-wave pairing cannot explain or solely accommodate for the large critical fields observed now by several groups. Recently, theoretical work by Lebed and Su [11] suggests the possible formation of Fulde-Ferrell-Larkin-Ovchinnikov (FFLO) [12,13] phase for magnetic field parallel to the conducting chains (*x*-axis).

Theoretically, it has been shown that orbital suppression of superconductivity in Q1D systems can also be reduced or eliminated by a field-induced-dimensional-crossover (FIDC) [14,15] effect, where an in-plane magnetic field reduces interlayer electron/quasiparticle motion, squeezing carrier motion to single layers and effectively decoupling the planes. As a result, magnetic field penetration (vortex cores) is restricted to the normal region between the planes. If this orbital suppression is accompanied by spin triplet Cooper pairing, which renders moot the Zeeman effect and the Pauli limit, then superconductivity can exist in arbitrarily strong fields. Alternatively, critical fields can exceed the Pauli paramagnetic limit for spin singlet or triplet pairing by the presence of a FFLO inhomogeneous state where a non-zero total momentum Cooper pair forms. The possibility remains, therefore, that superconductivity in $(TMTSF)_2X$ is initially spin singlet (*i.e.* at low magnetic field), and enters a spin triplet or FFLO spin singlet state in high magnetic fields. In either case, the anomalous $H_{c2}$ appears to be related to the Q1D nature, and so might be expected to be seen in other Q1D materials. Another potentially important consideration is the fact that the $(TMTSF)_2X$ materials which exhibit signatures of unconventional superconductivity are all in close competition with an antiferromagnetic SDW state, a situation that can be tuned with chemical or physical pressure. In fact, there is some evidence that when tuned far from the SDW-superconductor boundary, anomalous behavior in $H_{c2}$ diminishes [16]. This may suggest a role for spin fluctuations in the pairing mechanism in these materials. In this paper, we report the resistively-determined upper critical magnetic field of a closely-related Q1D organic superconductor, $(DMET)_2I_3$, which has strong structural and electronic ground state similarities with $(TMTSF)_2X$. Like $(TMTSF)_2PF_6$, it is free from any order-disorder transitions, and like $(TMTSF)_2ClO_4$, it superconducts at ambient pressure. Like both, it is susceptible to a field-induced spin density wave (FISDW) instability in a strong interlayer magnetic field but, unlike both, it does not undergo an SDW transition in zero field, and so appears not to be as proximate to the SDW-metal-superconductor phase boundary as those other materials exhibiting anomalous superconducting behavior [17].

**II. EXPERIMENT**

(DMET)$_2$I$_3$ [ 18 ] (dimethyl(ethylenedithio)diselenadithiafulvalene tri-iodine), a non-centrosymmetric molecular organic charge transfer salt with triclinic crystal symmetry, shows a superconducting transition at $T_c$ = 0.58K. As mentioned, it shares many similarities with the TMTSF system, including Q1D angular effects [19], superconductivity (though with only about half the $T_c$), electronic and triclinic crystal anisotropy, and FISDW transitions that are associated with its Q1D nature. It is comprised of highly conducting (Q1D) chains of DMET molecules along the crystal $b$ axis, organized into conducting layers along the $a$-$b$ plane that are separated along the $c$-axis by an I$_3$ anion layer. This Q1D crystal structure yields a pair of open Fermi surface sheets along the $k_a$-$k_{c*}$ plane, while the 0.5 electron per donor molecule charge transfer results in a 1/2-filled conduction band. In a Cartesian coordinate system, the orthogonal set ($x$, $y$, $z$) is represented by ($b$, $a'$, $c^*$), based on the lattice parameters $b$, $a$, and $c$. Interlayer resistance $R_{zz}(\theta,\phi)$ was measured on two samples, each with dimensions ~0.5 × 0.3 × 0.15 mm$^3$, using a dilution refrigerator and split-coil superconducting magnet equipped with a two axis ($\theta$, $\phi$) rotator, employed for the precise orientation of the samples with respect to magnetic field direction. The cryostat is situated on a goniometer which provides *ex situ* $\theta$–rotation (± 360º) with 0.05º resolution, and the samples were mounted on an *in situ* $\phi$–rotating (± 360º) platform controlled by a stepper motor-driven Kevlar string (also with precision ~ 0.05º). For field aligned along the $x$ and $y$ axes, the alignment of the sample accuracy is about ± 0.1º with respect to the $z$-axis.

## III. RESULTS AND DISCUSSIONS

Accurate crystal orientation with respect to the magnetic field direction has been shown to be essential to accurately determination of the upper critical fields in such highly anisotropic superconductors [3]. To determine the orientations of the orthogonal directions *x, y,* and *z* of the sample, we first measured $R_{zz}(\theta)_{\phi=90}$ in the normal state for magnetic field in the *y-z* plane. Figure 1 shows the angle dependent magnetoresistance for fields of strength 3, 6 and 9 T at 100 mK, rotating in this plane. The magnetoresistance has several oscillations whose angular positions can be indexed by the geometry of the crystal structure, the so-called Lebed magic angle effect [20]. We have also calculated [19] the magnetoresistance for the same field strengths and angular orientations for $(DMET)_2I_3$ using the triclinic lattice parameters [21] in the Boltzmann transport equation within the relaxation time approximation, as shown in Fig. 1. Curiously, the calculated magnetoresistance, while generally reproducing the angular oscillation effects, deviates significantly from the measured results for magnetic field in the vicinity of the *y*-axis ($\theta = 90º$). That is, the measured magnetoresistance has an anomalous, broad local minimum at $\theta = 90º$, while the overall tendency of the calculated result is toward a maximum. Furthermore, this experimentally-observed minimum deepens at higher magnetic field.

Similar behavior (minimum at along the *y*-axis in the normal state) has been observed in the TMTSF system [22, 23, 24], while all available theoretical models for interlayer magnetoresistance obtain a maximum. Classically in this orientation, the electron experiences a maximum Lorentz force, since $J \perp B$ there, so that a maximum in magnetoresistance is expected (and simulated) instead of the experimentally-observed minimum. Also note that the calculations in Fig. 1 yield a maximum for field not along *y*//*a'*, the normal to the planes at $\theta = 90º$, but at $\theta \sim 81.5º$, corresponding to reciprocal lattice $a^*$ direction. This suggests that the internal current flows along the intermolecular *c*-axis, as opposed to the $c^*//z$-axis anticipated by the macroscopic contact arrangement. This seems to be supported by the data: the 3 T experiment curve in Fig. 1, where the anomalous minimum is only weakly developed, exhibits a maximum near $a^*$, as seen in the right inset. This *H*//*y* direction is precisely that for which $H_{c2}$ is anomalously large in the (TMTSF)$_2$X system, so it might be tempting to associate the resistance decrease with superconductivity, via FIDC and/or FFLO [11, 25, 26]. Alternatively, Strong *et al.* have suggested that a large enough field parallel to this *y*-axis de-emphasizes coherent interlayer motion, transforming a 3D Fermi liquid into a 2D non-Fermi liquid [27]. If this effect is related to superconductivity, one might expect it to manifest itself on the upper critical fields. We have thus measured the upper critical field for magnetic field parallel to *y*-axis in (DMET)$_2$I$_3$.

Magnetoresistance of the sample itself can be used to orient the field *in situ* with respect to crystalline axes. As we have seen, $R_{zz}$ has an anomalous local minimum at the *y*-axis when the field is rotated from *z*. Rotating within the *x-y* plane next allows for precise alignment along the *y*-direction. As shown in Fig. 2, $R_{zz}(\phi)$ shows a local maximum centered at the *y*-axis for this rotation. In this rotation plane, current is ostensibly always perpendicular to magnetic field, so the Lorentz force is constant, and the angle dependence is controlled by Fermi velocity variation on the Q1D Fermi surface. As a result, the magnetoresistance has local minimum at *y* for a *y-z* rotation and a local maximum for an *x-y* rotation. The intersection of these two curves accurately defines the *y*-axis. Successive rotations of $\phi$ and $\theta$ by 90º give the magnetic field orientations parallel to the *x* and *z*-axes, respectively. With orientation fixed, four probe resistance versus magnetic field was measured at several temperatures. The RMS current used in the measurements was 1 µA, corresponding to $10^{-4}$ A/cm$^2$ density. Measurements were also carried out for higher (5 µA) and lower (0.1 µA) currents at the lowest temperature employed, 50 mK, to exclude artifacts from self-heating. Figure 3(a) shows resistance at several temperatures for field parallel to the *y*-axis. These plotted data are the interpolated points after the curves were digitally smoothed using data analysis software. Similar measurements were carried out for fields parallel to the *x* and *z*- axes.

The transition out of the superconducting state, upon increasing the magnetic field, is seen as a gradual rise of resistance, ending in a quasi-linear dependence on field in the normal metal state. Since there is no well-defined rule to extract $H_{c2}$ from such experiments, we employed the criteria shown in Fig. 3, demonstrated by repotting the 50 mK data after shifting the field by 0.05 T: O (onset), J (junction), M (midpoint), X (extrapolation to zero resistance) and Z (zero resistance). The resulting phase lines are plotted in Fig. 4 for this orientation, $H//y$. All the curves show similar temperature dependencies, including a slight slope change (toward more rapid suppression of superconductivity) upon cooling near $T_c/2$, with an even more subtle indication of positive curvature for the lowest field curves below 0.1 K. We plot in Fig. 5 the upper critical field versus temperature from Fig. 4 and from similar data for $H // x$ and $z$ using the M criterion. In addition, for some upper critical field datum points, the sample was cooled to base temperature in constant field along the *x*-axis. The error bars are comparable to the size of the data points in the figure.

$H_{c2}$ in Fig. 5 tends to saturation at low temperature for field along all three principal axes, with $H_{c2}^x > H_{c2}^y > H_{c2}^z$, as expected for a Q1D superconductor from the normal state anisotropy. We also plot fits to these data using the Werthamer-Helfand-Hohenberg (WHH) formula [28], which seem to reproduce the observed $H_{c2}$ temperature dependence reasonably well for all three directions, $H_{c2}(T)$ saturating as $T$ approaches zero. We have extracted from these fits estimated values of the zero temperature critical field along *x*, *y*, and *z* as 0.79 T, 0.186 T and 0.019 T, respectively. Using the Ginzburg-Landau relation, $H_{c2}^{i}(0)=\phi_o/2\pi\xi_j(0)\xi_k(0)$, where $\phi_0$ is the flux quantum and $\xi_i(0)$ is the zero temperature coherence length along the $i^{th}$ direction, these anisotropic coherence lengths can be obtained as 271 nm, 64 nm, and 6.5 nm, along *x*, *y*, and *z*, respectively. That is, the anisotropy of the coherence length is found to be $\xi_x : \xi_y : \xi_z$ = 41.2 : 9.8 : 1. On the basis of the tight binding approximation, this anisotropy, due to orbital effects, is related to the band structure anisotropy via $\xi_x : \xi_y : \xi_z = (a_x/2)t_x : a_y t_y : a_z t_z$, where $a_i$ and $t_i$ are the lattice parameters and transfer integrals, respectively [29]. Using the lattice parameters of $(DMET)_2I_3$ and the calculated coherence length anisotropy, we estimate the transfer integral anisotropy to be $t_x : t_y : t_z$ = 194 : 20 : 1. The same anisotropy ratio of $t_x/t_y = 9.7$ was found in this material from the Yoshino angular effect measurements [30].

As mentioned, there are two pair breaking effects of magnetic field in superconductivity, of orbital and spin origin. When the interlayer coherence length in a Q1D superconductor is comparable to the interplane distance, *i.e.*, $\xi_z(0) \sim c^*$ at sufficient high magnetic field, FIDC can suppress orbital pair breaking and allow superconductivity to persist. This may be the case in $(TMTSF)_2X$ (X = $ClO_4$ and $PF_6$). However, in $(DMET)_2I_3$, the interplane coherence length $\xi_z(0)$ is about four times the interlayer distance 1.55 nm, suggesting that FIDC would not occur. On the other hand, the spin pairbreaking Pauli limit in a 3D system is given by $H_P(T=0) = 1.84T_c(H=0)$ for isotropic *s*-wave pairing in the absence of strong spin-orbit coupling [31,32], or $1.58T_c(H=0)$ for the case of anisotropic (2D) singlet pairing [33]. Using the observed superconducting transition temperature, $T_c$ = 0.58 K, $H_P$ should fall between 0.9 T and 1 T. The observed $H_{c2}$ along all three directions is smaller than these calculated values of $H_p$, as shown in Fig. 5. Therefore, the upper critical field in $(DMET)_2I_3$ does *not* exceed the paramagnetic limit, in contrast to the case of $(TMTSF)_2X$. Thus, the anisotropy in the upper critical field is due to orbital pair breaking, linked to the band structure anisotropy. The nature of superconductivity in (DMET)$_2$I$_3$, from the viewpoint of $H_{c2}$, appears to be conventional, quite unlike the isostructural Bechgaard salts.

Based on our results, the observed magnetoresistance minimum in the normal state (Fig. 1) is not associated with superconductivity. A similar minimum has been observed in the non-superconducting Bechgaard salt (TMTSF)$_2$NO$_3$ for field along the *y*-axis [34], though this compound is believed to be a semimetal with a Q2D Fermi surface at low temperature [35]. A possible reason for the minimum in the present (DMET)$_2$I$_3$ case could be a type of normal state FIDC, where the amplitude of electron trajectories become comparable to the interplane distance. This amplitude is given by $2a_z t_c / \hbar \omega_c$, where $\hbar \omega_c = e H a_z v_F$ ($v_F$ is the Fermi velocity). With the typical values $t_c \sim 1$ K, $a_z = 1.58$ nm and $v_F = 4.0 \times 10^4$ ms$^{-1}$ [36], the field at which the FIDC would occur is 2.7 T. This threshold field for FIDC is higher than the superconducting critical field, but it is comparable to the field regime where the anomalous minimum develops (~3T, Fig. 1 inset).

## IV. CONCLUSION

We have measured the resistive upper critical field of (DMET)$_2$I$_3$ for field along the *x*, *y*, and *z*-axes, and found that it tends to saturation at low temperature for all directions, following the conventional WHH-formula. This is unlike the unconventional, non-saturating behavior observed in the (TMTSF)$_2$X Bechgaard salts, thought to be associated with spin fluctuation-driven pairing. However, with the reduced $T_c$ (and thus $H_{c2}$) relative to (TMTSF)$_2$X, the results may be consistent with the case for conventional Q1D superconductivity at low field (below ~1 T) and an unconventional state at high field (driven by charge and/or spin fluctuations) for those systems which survive beyond 1 T [9,10,37]. (DMET)$_2$I$_3$ may just have a $T_c$ too low to reach this latter condition, being too far from the SDW instability on the generalized phase diagram. The minimum observed in magnetoresistance is associated with a normal state FIDC, rather than superconductivity.

## V. ACKNOWLEDGEMENTS

This work was supported by the National Science Foundation, under Grant No. DMR-0605339. *Present address: Jefferson Lab, Newport News, VA 23606 USA. +naughton@bc.edu

Figure captions

FIG. 1. (color online) Angle-dependent magnetoresistance of $(DMET)_2I_3$ at 3, 6 and 9 T, rotated in the *y-z* plane ($\phi = 90°$), measured at 100 mK (solid lines) and calculated using triclinic Boltzmann transport equation (dashed lines). Right inset: Expanded view of 3 T results, showing anomalous minimum developing at *H//a'* in experiment. Left inset: Axis and angle definitions.

FIG. 2 (color online) Magnetoresistance at 6 T and 100 mK used to precisely locate the sample *y*-axis, via field rotations in the *y-z* (open circles) and in the *x-y* planes (solid circles).

FIG. 3 (color online) Magnetic field dependence of interlayer resistance of $(DMET)_2I_3$, normalized to the normal state value $\rho_N$, for field parallel to the *y*-axis at different temperatures. The lowest temperature curve (50 mK) is replotted with *H* shifted by 0.05T to indicate various criteria employed to determine $H_{c2}(T)$ from $T_{c2}(H)$: O (onset), U (junction), M (midpoint), X (extrapolation to *R*=0) and Z (zero resistance).

FIG. 4 (color online) *H-T* phase diagram for superconducting state of $(DMET)_2I_3$ for field aligned along the *y*–axis, for $H_{c2}$ criteria identified in Fig. 3. Lines are guides to the eye.

FIG. 5 *H-T* phase diagram in $(DMET)_2I_3$ for field aligned along *x*, *y*, and *z* axes, using the midpoint criterion. Lines are calculated using the WHH formula. Pauli limits for isotropic (3D) and anisotropic (2D) *s*-wave pairing in the absence of strong spin-orbit coupling are indicated.

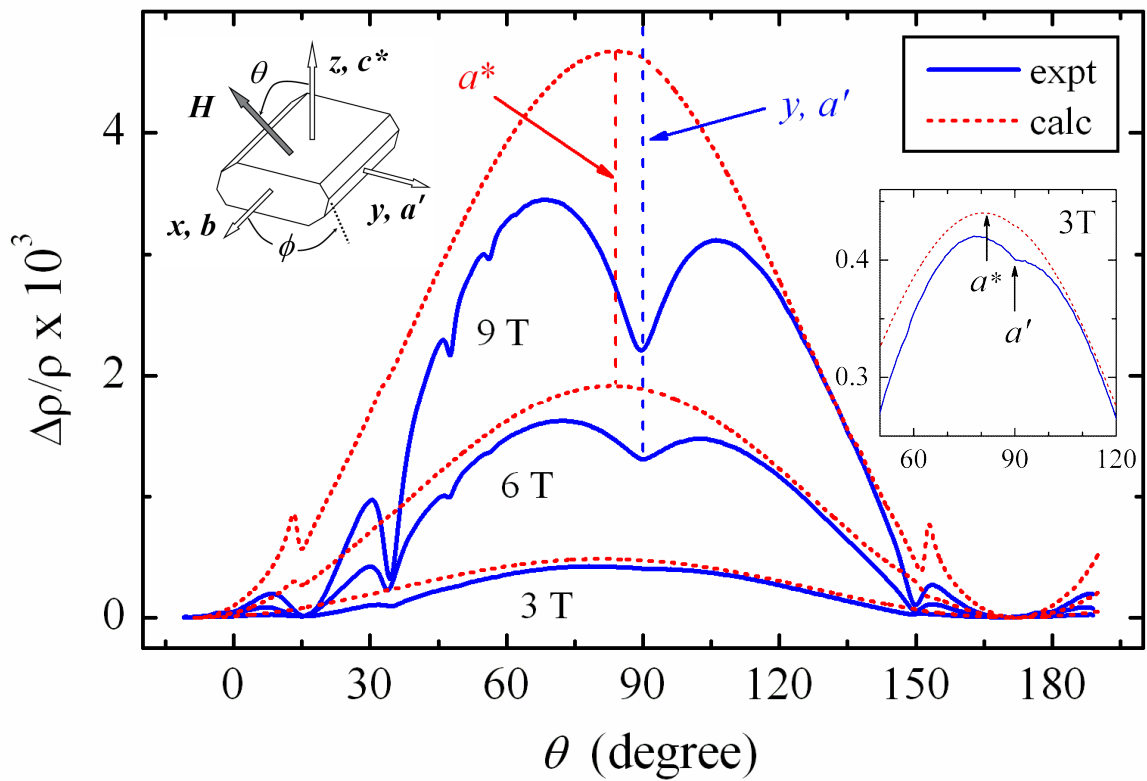

FIG. 1

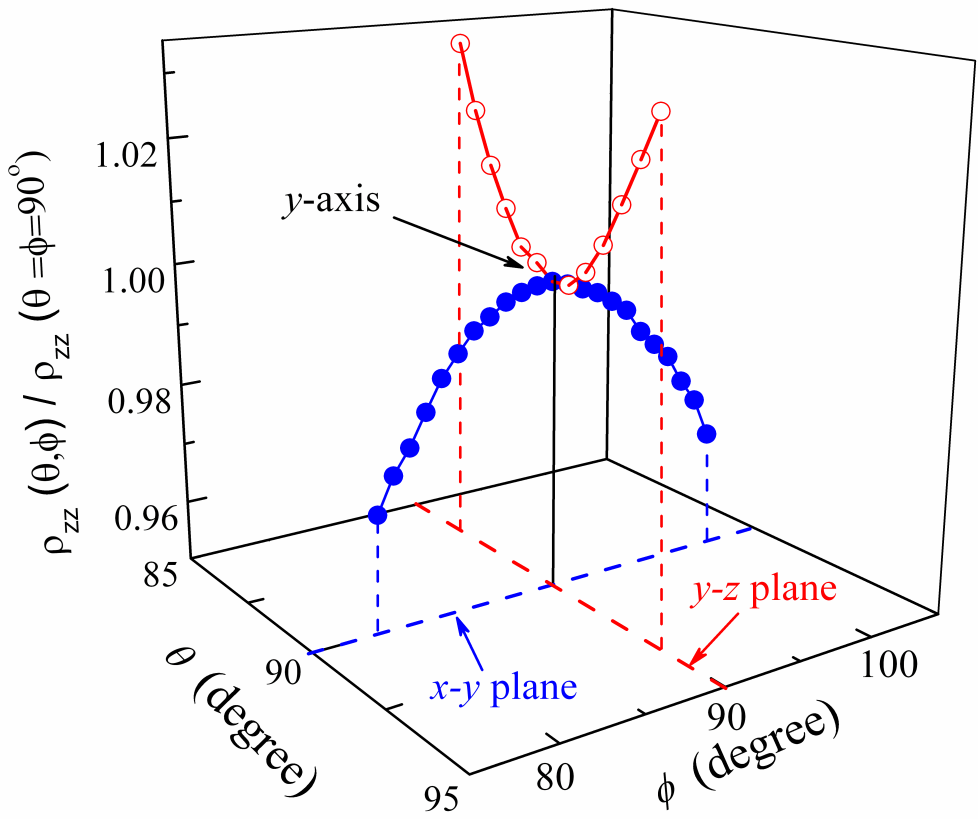

FIG. 2

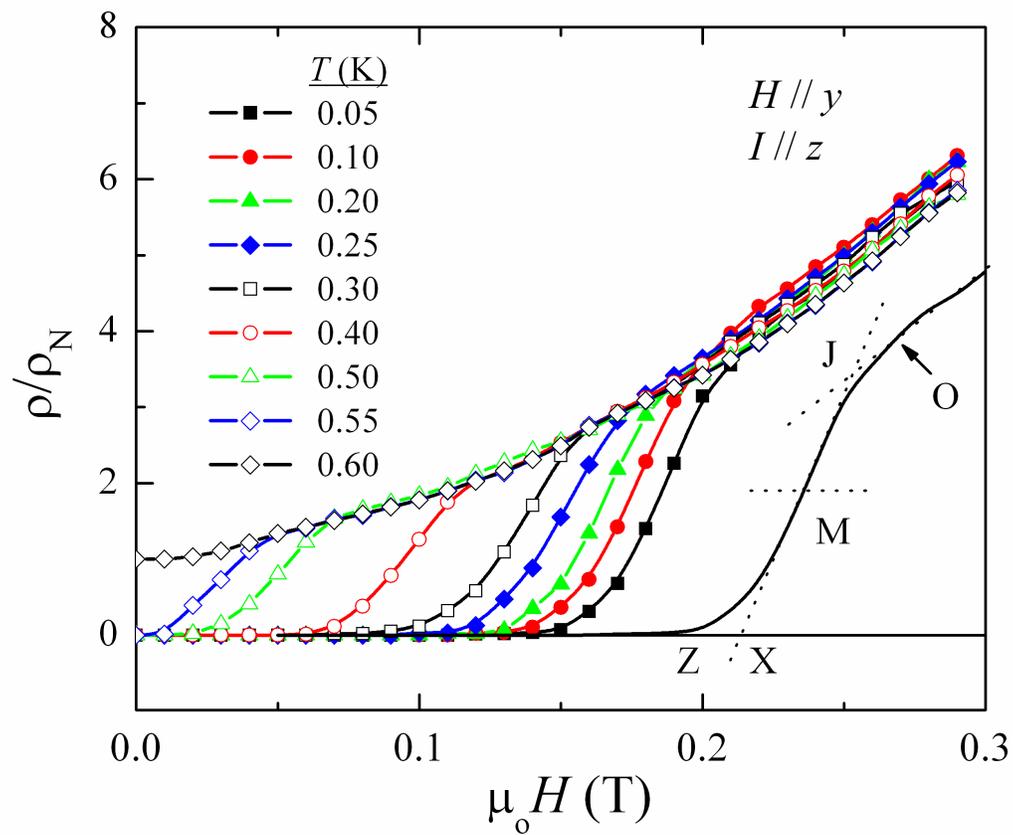

FIG 3

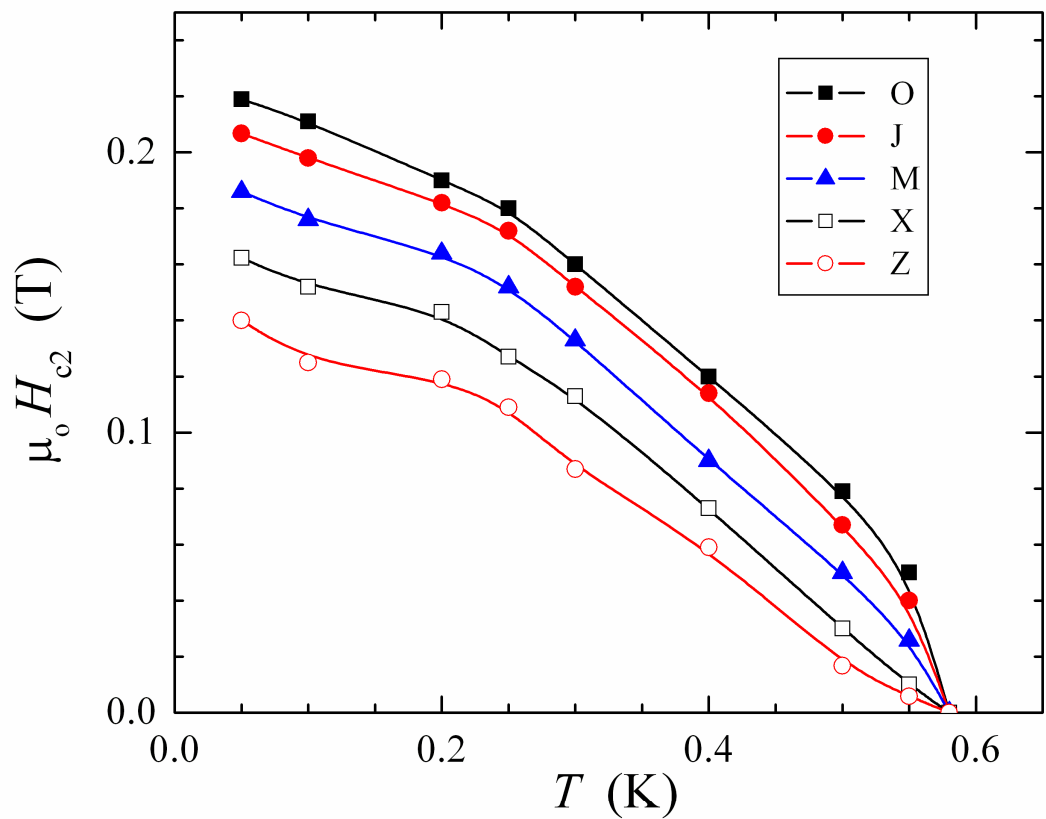

FIG. 4

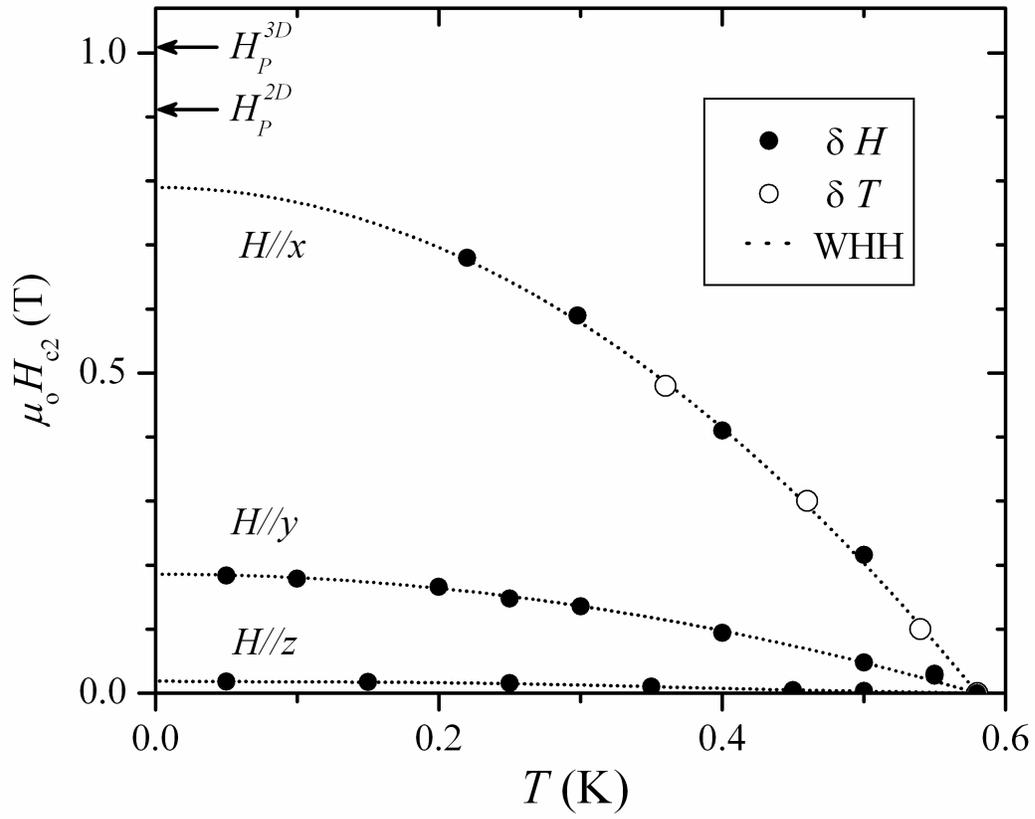

FIG. 5